\documentclass[10pt, conference, letterpaper]{IEEEtran}

\usepackage{amsmath}
\usepackage{cases}
\usepackage{graphicx}

\usepackage{cite}
\usepackage{times}
\usepackage{graphicx}
\usepackage{graphics}
\usepackage{subfigure}
\usepackage{epsfig}
\usepackage{latexsym}
\usepackage{amsfonts}
\usepackage{paralist}
\usepackage{xspace}
\usepackage{mathrsfs}
\usepackage{amssymb}
\usepackage{color}
\usepackage{algorithm}
\usepackage{algorithmic}
\usepackage{mathtools}

\usepackage{amssymb}
\usepackage{url,epsfig,array}
\usepackage{tikz}
\usepackage{caption}

\allowdisplaybreaks %%% this make pages look tight.
\def\ie{\textit{i.e.}\xspace}

\def\etc{\textit{etc.}\xspace}

\def\eg{\textit{e.g.}\xspace}

\def\path{{\textbf{P}}}
\newcommand{\CUTXY}[1]{{}}

\newcommand{\dis}{\mathsf{d}}
\newcommand{\Sim}{\mathbf{S}}
\newcommand{\vx}{\mathbf{x}}
\newcommand{\vy}{\mathbf{y}}
\newcommand{\va}{\mathbf{a}}
\newcommand{\vv}{\mathbf{V}}
\newcommand{\code}{\mathbf{SC}}

\newcommand{\lsh}{\mathbf{H}}

\renewcommand{\paragraph}[1]{\smallskip \noindent {\textbf{#1}}}

\def\ourprotocol{\text{InvisibleMe\ }}
\def\basic{\text{InvisibleMe-Basic\ }}
\def\advanced{\text{InvisibleMe-Advanced\ }}

\begin{document}
%\title{SeeNet: }
\title{Enable Portrait Privacy Protection in Photo Capturing and Sharing}

%\author{IEEE INFOCOM 2015, Paper ID: 398}

\author{\authorblockN{Lan Zhang\authorrefmark{1}, 
Kebin Liu \authorrefmark{1},
Xiang-Yang Li\authorrefmark{2},
Puchun Feng\authorrefmark{1},
Cihang Liu\authorrefmark{1},
Yunhao Liu\authorrefmark{1}}
\authorblockA{\authorrefmark{1} School of Software, Tsinghua University}
\authorblockA{\authorrefmark{2} Department of Computer Science, Illinois Institute of Technology}
}

\maketitle

\begin{abstract}
The wide adoption of wearable smart devices with onboard cameras
 greatly increases people's concern on privacy infringement.
Here we explore the possibility of easing persons from photos captured by
 smart devices according to their privacy protection requirements.
To make this work, we need to address two challenges: 1) how to let
 users explicitly express their privacy protection intention, and 2)
 how to associate the privacy requirements with persons in captured photos accurately and efficiently.
Furthermore, the association process itself should not cause portrait information leakage
 and should be accomplished in a privacy-preserving way.
In this work, we design, develop, and evaluate a protocol, called
\emph{InvisibleMe}, that enables
a user to flexibly express her privacy requirement and empowers the
 photo service provider (or image taker) to exert the privacy
 protection policy.
Leveraging the visual distinguishability of people in the field-of-view
 and the dimension-order-independent property of vector similarity measurement,
 InvisibleMe achieves high accuracy and low overhead.
%As image processing is expensive (in terms of computation and
% communication cost), we carefully design our protocol by leveraging
% the power of cloud computing.
We implement a prototype system, and our evaluation results on both
 the trace-driven and real-life experiments confirm
  the feasibility and efficiency of our system.
\end{abstract}

%\keywords{Photo Sharing, Portraiture Privacy, Wearable Smart Camera}

\section{Introduction}
\label{sec:introduction}
Nowadays, smart devices with onboard cameras  \eg, smart phones and
glasses, are pervasive in our daily lives.
These smart devices can capture and even share photos without
 informing the  parties in the picture, thus raises many
 concerns on people's  privacy infringement.
Particularly, the wilder adoption of smart glasses, \eg Google Glass,
 leads to severe concerns for misuse because Glass can capture
 photos/videos far less conspicuously than a traditional hand-held
 device.
%For example, you cannot see whether a Google Glass is recording and there is no warning of it.
Secretive photographing without clear warning beforehand and
 possession of secretively taken photos are both privacy violations.
Even worse, if the photos which contain information beyond what
 users want to reveal are shared in Internet, it will make users
 extremely susceptible to various attacks.
% \cite{zhou2008preserving}.

To protect people's portrait privacy from unwilling photo-taking and publication,
 many photo service providers or users have taken actions in different ways.
For example, some Glass wearers whip their device off in inappropriate situations,
 such as in gym locker rooms or work meetings;
 some business bans smart glasses inside their buildings to respect
 customers' privacy~\cite{banglass}; and
 the Glass manufacturer (\eg, Google) does not allow developers to
 create applications that take photo silently.
%But they remain silent on whether they reserve themselves the right to do it.
Lawmakers are also beginning to consider various privacy issues of
 Glass, including whether it should be capable of facial recognition \cite{google-law}.
Although face recognition is a useful function, especially for social applications,
 face is a critical private identifiable information.
At present, there are no facial recognition technologies built into Glass
 and the manufacturer has no plans to use it unless they have strong privacy protections in place.
%Social implications and etiquette have been a big area of focus during the development of the camera product,
These methods, however, are broad-brush and blunt which can
 significantly hurt the applications of smart devices.
Therefore, it is appealing to consider how one might build a system in
 which users do not leak portrait privacy while guaranteeing a
 comfortable usage of smart glasses/cameras.

Instead of discarding the smart glasses/cameras due to privacy concerns,
 in this work, we seek a solution for reaching an ultimate goal of
 privacy-friendly Glass/camera, operating  transparently to
 end users.
Our solution will let end users to express their privacy requirements
 and glasses/cameras or photo service providers will exert the privacy
 protection mechanisms.
When taking a photo/video, the smart device will detect who (in the
picture/video) requested privacy protection, and then remove them
 from the image automatically.
Our protocol can also be used for automatically tagging people in a photo
%\cite{qin2011tagsense}
when a
user expresses an interest to be tagged with his/her information.

To implement such a privacy-friendly camera, we need to address
several critical challenges.
\textbf{First}, we should enable privacy-advocator efficiently and flexibly to express his/her privacy
requirements/intentions. Several methods could be used for this task, such as
using visible specialized tag (\eg, QR code), or encoding the request and transmitting
it using  wireless devices.
For users' convenience and aesthetics, in this work, we adopt the latter approach by encoding his/her
portrait in the request.
Then, the \textbf{second} challenge is that we should accurately and efficiently associate
 each privacy-seeking user with an image region in the photo taken by another user (the photographer ).
Furthermore, the association process itself should not cause portrait information leakage
 and should be accomplished in a privacy-preserving way.
Face recognition \cite{luo2007person, turk1991eigenfaces} is
widely used to identify people in photos, but in practice it suffers when there lacks a clear front view of faces.
%due to the camera's view angle and distance.
%For example, in our field studies, when there are 1326 pedestrians detected, only 412 faces are recognized.
Sophisticated but complicated matching schemes may cause high
overhead and long delay. The matching problem itself is difficult due to
the accuracy and efficiency requirements, let alone completing
matching process in a private and non-interactive manner with untrusted server.
Matching a user's privacy-expression with a possible people in a photo
can be reduced to some sort of vector matching.
% or dot multiplication operation.
 Many private vector matching protocols use
multi-party computation techniques, which require frequent
interactions among participants. Most exiting private vector matching
methods (in both multi-party computation and outsourced manner) use
homomorphic encryption \cite{katz2008predicate, sadeghi2010efficient} or
garble circuit \cite{sadeghi2010efficient},
and cause high computation cost for both client and cloud.
The \textbf{third} challenge is that the privacy-friendly Glass/camera should be
 transparent to all users and cause minimal extra overhead to mobile devices.
%Obviously, if the proposed solution distracts the user experience
% while using their smart devices, they may decline to install the
% functionality.
An ad hoc approach may lead to requirements for "always-on" neighbour
 discovery, frequent information exchanging as well as complex image matching
 computation on user devices.
To reduce the overhead of users,
 our protocol will outsource most of these tasks to cloud with a
 well-designed strategy to prevent privacy leakage to untrusted cloud and other users.
%In this work, we apply the widely used assumption about the cloud to
% be "honest-but-curious" which keeps on trying to infer information
% from users while follows the protocol.

The main contributions of this work are as follows:

%\begin{compactenum}
\textbullet To the best of our knowledge, we are the first to present a
  portrait privacy preserving photo capturing and sharing approach.
  % which enables a user to express
  %his/her privacy requirement and empowers the smart devices to exert the protection.
 With our approach in Section~\ref{sec:design}, people who require not to be captured in photo will be
 automatically erased from the photo and verification of the removal is also supported.

\textbullet We comprehensively analyze the privacy issues during the photo
  capturing and sharing and define three types of threats in Section~\ref{sec:system_model}. Based on
  the proposed model, we present a solution protecting all three types of privacy information.

\textbullet For accurate and efficient matching between people's privacy
  intentions and people in the photo,
  we introduce a graph-based portrait profile and design a robust
  matching algorithm to recognize and erase privacy-seeking people in Section~\ref{sec:matching}.

\textbullet  We propose a highly efficient privacy-preserving vector distance
  protocol in a non-interactive manner with untrusted server in Section~\ref{sec:advanced},
  which significantly reduces the computation complexity and communication cost than
  existing homomorphic encryption and garble circuit based solutions.
%  our approach leverages the dimension-order-independent property of distance
%  between vectors and the locality sensitive hashing to enable the distance computation on transformed vectors other than cypher blocks,
%  which significantly reduces the computation complexity and communication cost.
  With our protocol, most computation tasks are transferred from smart devices to the cloud in a privacy-preserving way.
  %We present the approach with advanced privacy preserving guarantee, which protects users' portrait privacy while
  %transfers the computation tasks from users' smart devices to the cloud in a privacy-preserving way.

\textbullet We design and implement a prototype system and verify the effectiveness of our scheme by extensive experiments as well as case studies in Section~\ref{sec:implementation}.
%\end{compactenum}

%The rest of this paper is organized as follows.
%First we introduce the motivation and system model in Section~\ref{sec:system_model}.
%Section~\ref{sec:design} presents the overview of this work.
%We give the graph-based portrait matching algorithm in Section~\ref{sec:matching} and
% the privacy-preserving vector distance protocol in Section~\ref{sec:advanced}.
%We discuss the implementation of our prototype system and evaluate the effectiveness of our system
% with real life experiment in Section~\ref{sec:implementation}.
%Related work is discussed in Section~\ref{sec:related}.
%We conclude this work in Section~\ref{sec:conclusion}.

\vspace{-0.05in}
\section{Privacy Requirement}
\vspace{-0.08in}
\label{sec:system_model}
\subsection{Motivation}
\vspace{-0.05in}

To protect users' portrait privacy,
one straight-forward approach that has already been adopted by
business and manufacturer is to simply suppress the usage of
smart cameras in specific place and time, \eg, turn off the
photographing functionality in a meeting room or forbid silent photo
taking.  These
methods, however, are broad-brush and blunt which can significantly
hurt the applications of smart devices. Besides, they do not meet the
user varying privacy intentions. In fact, not all persons are
unwilling to be photographed and also people will feel quite
uncomfortable if they are frequently forced to turn off their Glass.
% or
%obtain pictures with all faces blurred.
Thus, in this work we seek solution to give privacy control back to
persons being photographed in a smarter way.

\begin{figure}[t!]
\begin{center}
\includegraphics[width=0.8\linewidth, clip,keepaspectratio]{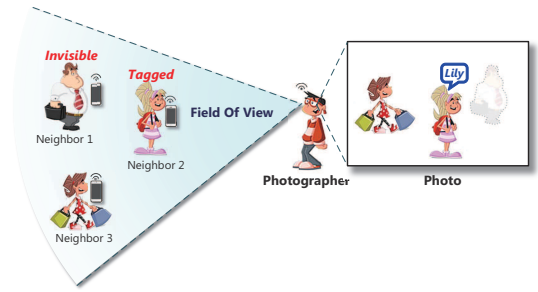}
\caption{Example application scenario of \emph{InvisibleMe}: the invisible person is erased from the photo and the tagged person is labeled in the photo.}
\label{fig:application_model}
\end{center}
\vspace{-0.3in}
\end{figure}

As an example shown in Fig.~\ref{fig:application_model},
when someone uses his smart Glass to take a photo, people in the field
of view (FOV) should be notified (or the photographer should know the
privacy protection intentions of people in FOV).
Then persons who are unwilling to be photographed, \eg Neighbor 1,
should have a convenient way to specify their privacy intentions, and
thus be automatically erased from the photo.
We refer to them as \emph{invisible users}.
Users that would like to make friends with the photographer, \eg, Neighbor 2,
can be automatically tagged on the photo and share information.
We refer to them as \emph{tagged users}.
The photographer just takes other people into the photo as usual.
The system can motivate the photographer by mutual registry:
 only those who protect other invisible users can be registered to be protected by others.
Supporting tagging people automatically, which could be helpful and fun in many scenarios (\eg, facebook),
 also gives incentives to the photographer.
Besides, after one time setting the solution should be transparent to
all users and avoid incurring high overhead to their smart devices.
Finally, the portrait privacy protection strategy should be well
designed and avoid further leakage of any type of private information.

\vspace{-0.08in}
\subsection{Threats to Portrait Privacy}
\label{sec:threat}
\vspace{-0.05in}
%Photo taking is available for most smart devices(\eg, mobile phone,
%ipads and Glass), and photo sharing services (\eg, Instagram and
%Facebook) are getting extremely popular.
%People may be photographed nearly anytime anywhere without getting
%their consent.
%For example, one cannot notice that a Glass is taking a photo of him/her.
Photos contain rich information, including people's appearance, location, activities, \etc.
%There are many mature techniques to detect and recognize the objects (\eg, faces and pedestrians)
% within the photos\cite{leibe2005pedestrian, turk1991eigenfaces}.
%Many image analysis techniques, \eg, \cite{leibe2005pedestrian} and \cite{turk1991eigenfaces}
% can possibly be used to automatically mine sensitive information from photos.
Facing massive cameras and image analysis techniques \cite{leibe2005pedestrian, turk1991eigenfaces} , people's portrait privacy is badly in need of
protection.
In this work, we focus on protecting users' portrait information.
Here, a portrait not only includes the user's face but also his/her body,
 since clothes and accessories could also reveal identification information.
We consider three types of threats to portrait privacy.
%, which expose sensitive information to different extend.

\paragraph{Visual portrait privacy.}
 The most intuitive way to violate a user's portrait privacy
 is to capture and publish (\eg, through photo sharing systems) a photo containing his/her visible portrait.
Simply blurring all faces in images, \eg, \cite{simoens2013scalable} and Google Street View,
% cannot be applied to photo capturing of personal device because
 will disable the normal photographing function.

In our protocol \ourprotocol we propose to match the people in the
photo to their privacy protection intentions,
 and erase only  people that should be invisible.
%There are several critical steps in designing a privacy respecting
%photo-taking and sharing system, including, but are not limited to,
%1) privacy expression by potential users in the photo, and
%2) privacy respecting mechanism for the photographer.
As we will discuss in detail in Section~\ref{sec:design},
a user express his/her privacy requirement by encoding his/her portrait,
which clearly cannot be transmitted in its original form
(otherwise his/her portrait privacy is broken by himself/herself).
%Receiving such privacy expression, the photographer need to
%perform matching to find the correct person to erase.
So, we need to provide privacy protections in all these operations.

\paragraph{Portrait feature privacy.}
This type of threats occur inside some image services, \eg
 image matching or face recognition.
These services don't use visible images of directly, but take feature
 vectors of image as the descriptor, \eg, Eigenfaces \cite{turk1991eigenfaces} and color histogram.
%Various feature vectors, \eg, Eigenfaces \cite{turk1991eigenfaces},
%feature point descriptors and color histogram,
%could be used for image matching and recognition.
% \eg color and face descriptors \cite{turk1991eigenfaces}.
But users can also be identified by features of portrait image.
For example, face images can be reconstructed from face vectors\cite{facereconstruction}.
During the process, the leakage of portrait features also violates users' privacy.

%Vision feature vectors, \eg, Eigenfaces \cite{turk1991eigenfaces}, feature point descriptors and color histogram,
% are widely used for object detection and recognition, and image retrieval.
%They don't give away visible portrait of people directly, so usually are ignored to be treated as privacy.
%However, an adversary can reconstruct the original image with feature vectors.
%For example, faces can be reconstructed from eigenfaces vectors. \mynote{citation here}
%We refer to the vision feature vectors of a portrait as portrait feature.
%Portrait feature is also an important part of portrait privacy.

\paragraph{Inference privacy.}
Even if an image system hides original images and other information such as their feature vectors,
 an adversary with a collection of images (an image dictionary) can infer the hidden content using the similarity measurement function
 of the system.
 Hence, we should prevent adversaries from obtaining the similarity measuring results to enhance the privacy protection.

%There are also other user privacy should be considered in the system design, \eg location privacy.
%A lot of methods have been proposed to provide privacy-preserving location services \cite{li2013search}.
\ourprotocol leverages existing solutions to protect other user privacy, \eg, location privacy \cite{xu2009feeling}, since it is not the focus of this study.

%%%%%%%%%%%%%%%%%%%%%%%%%%%%%%%%%%%%%%%%%%%%%%%%%%%%%%%%%%%%%%%%%%%%%%%%%%
\CUTXY{
\subsection{Challenges}

There are many challenges associated with the task.

First, how to let users efficiently and flexibly express his/her
privacy requirements or intentions.

Second how to accurately and efficiently associate each user that has privacy intention to an image region in a privacy preserving way.
Some existing efforts propose to tag people in image using face recognition \cite{bicego2006use,luo2007person, everingham2006hello, turk1991eigenfaces}. That is, users send their face images to adjacent photographer and the photographer erases or tags the matched people in the photos.
These methods, however, suffer when there lacks a clear front view of face due to the camera's view angle or weak illumination.
According to our field studies, when there are 1326 pedestrians detected, only 412 faces are recognized.
Besides, broadcasting face images leads to noteworthy privacy concern which is vulnerable to many attacks.
To address the above issues, in this work we will leverage both face
and other dimensions of human vision features such as body and
clothing color. In fact, these features are very efficient in distinguishing people in a local area.

Third, the protocol should respond to the privacy request accurately
within a reasonable delay while causing minimal services disruption to
end users.
Notice that a key step here is to match users in the photo with users
who requested privacy protection.
To achieve privacy preserving match, many existing protocols
use multi-party computation techniques, which require frequent
interactions among participants and apply homomorphic encryption or
garble circuit that cause high computation cost.
Instead, we introduce a graph based matching mechanism together with
Locality Sensitive Hashing and scramble vector, achieving highly
matching accuracy and computation efficiency while guaranteeing user
privacy.

Fourth, how to avoid introducing extra overhead on users as well as
 their devices.
In existing works, the people tagging task is accomplished using face
recognition techniques
 and proximity wireless communication, \eg Wi-Fi and Bluetooth.
In this case, however,
 the photographer needs to conduct neighbor discovery for every photo taking.
And devices of all invisible users are required to be on the standby mode to
 receive every  photographing notification.
Neighbor discovery and wireless communication establishment in ad hoc
 mode are complicated and also power-hungry,
which makes the ad hoc solution unpleasant for mobile users.
Besides, image matching could cause high computation overhead for the photographer and thus make photographers decline to use the functionality.
Using an always online powerful cloud could be a better choice to provide the location service, cache and relay messages
 as well as conduct all costly computation.
So, in our system, there are three parties involved: the photographer, who takes the photo;
the neighbors, who are in the FOV of the photographer;
the cloud, who takes charge of location, communication and computation services.
%The cloud could be any photo service provider, \eg Instagram and Facebook,
% who have the motivation to provide better privacy protection to attract more users.
In this case, we face with the critical challenge of leveraging the resources of cloud while preventing the cloud from inferring users's privacy.

}%end of cut
%%%%%%%%%%%%%%%%%%%%%%%%%%%%%%%%%%%%%%%%%%%%%%%%%%%%%%%%%%%%%%%%%%%%%%%%%%%%%%%%%%%%%%%%%%%%%%%%%%

\paragraph{Adversary  model.}
%As we have discussed above, there are three parties engaged in \ourprotocol:
%photographer, neighbors in FOV and a cloud server.
%Obviously,
% adversarial participants are more powerful than outsiders,
% therefore we only consider participating adversaries: adversarial
%cloud server and users.
Our approach defends a user's portrait privacy against both untrusted cloud server and malicious users.
For the cloud, we apply the widely used "honest-but-curious" assumption.
The cloud server will follow the protocol,
 but might conduct extra work to harvest portrait images of invisible people,
 reconstruct invisible portraits using feature vectors or infer the invisible content using image dictionary.
%This is a justifiable assumption because deviating from the protocol
% will lead to incorrect people erasing results
% and bad user experience could cause revenue loss of the service provider.
Also, we assume the cloud won't collude with any client to conduct an attack.
%All these are sensitive information requiring protection.
A malicious user could participate to harvest
 other users' portrait information by eavesdropping their communication with the cloud.
All users except the photographer should be prevented from obtaining the portrait information of invisible users.
Although the photographer already owns the photo of invisible people,
 he/she may misbehave to preserve the invisible people who should be erased and publish the photo through Internet.
So, we also need a verification scheme against dishonest photographers.

\vspace{-0.08in}
\section{System Design Overview}
\vspace{-0.08in}
\label{sec:design}
\begin{figure}[t!]
\begin{center}
\includegraphics[width=0.7\linewidth,height=2in,clip,keepaspectratio]{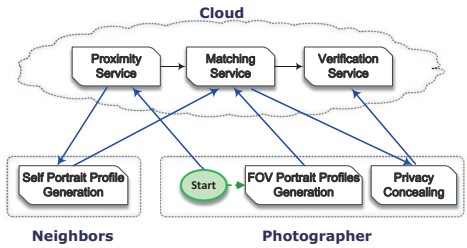}
\caption{Baseline system architecture.}
\label{fig:basic}
\end{center}
\vspace{-0.25in}
\end{figure}

%In this work, we firstly present our baseline approach based on graph matching, locating invisible people in images and protecting them from being photographed.
Facing critical challenges introduced in Section \ref{sec:introduction},
 we design our system to achieve both the privacy and system efficiency goals.
%We firstly present our baseline system design.
With our graph-based portrait matching algorithm, the baseline approach is effective to protect users' visual portrait privacy
 by accurately locating invisible people in photos and erases them automatically.
But there is a risk of exposing portrait features to untrusted cloud and other participants.
Furthermore, we propose an efficient privacy-preserving outsourced vector distance protocol,
 based on which an advanced approach provides portrait feature privacy and inference privacy protection with little extra overhead for the client.
In this section, we will sketch our system architecture.

\subsection{Overview of Baseline System}
\vspace{-0.05in}
In our system, there are three parties involved: the photographer, who takes the photo;
the neighbors, who could be in the FOV of the photographer (as presented in Fig.~\ref{fig:application_model});
the cloud, who takes charge of location, communication and computation services.
%In our system model, there are three group of basic participators, the photographer, the neighbors (users adjacent to photographer) and the cloud.
%The photographer takes pictures of surroundings and some neighbors will appear in the photo. As we have discussed, some neighbors want their portrait privacy to be protect and thus are not willing to be photographed.
%Thus, our goal is to remove the unwilling people from the photo before it is kept and shared.
The architecture and workflow of our baseline system are illustrated in Fig.~\ref{fig:basic}
%, our baseline system includes the following major components that are deployed on photographer, neighbors and cloud separately.
%\begin{compactitem}

%\item The \emph{Self Portrait Profile Generation} component, located at
%  the neighbor side, creates personal portrait profile of the user.
%  The user will encode his/her portrait profile to express his/her privacy requirement.

%\item The \emph{FOV Portrait Profiles Generation} component deployed on
%  the photographer extracts portrait profiles of people from the taken photo.

%\item The \emph{Proximity Service} on cloud helps to detect users in
%  proximity that may appear in photographer's FOV.
%This component can reduce the computation cost of cloud and communication cost of end users.

%\item The \emph{Matching Service} on cloud conducts portrait profile
%  matching tasks to determine people that should be erased from the photo.

%\item The \emph{Privacy Concealing} component, located on photographer,
% erases invisible people from the photo automatically before uploading.

%\item The \emph{Verification Service} runs on cloud side to take charge of verifying
% the removal of invisible people, in case there are dishonest photographers.
%\end{compactitem}

\begin{figure}[t!]
\begin{center}
\includegraphics[width=0.7\linewidth, clip,keepaspectratio]{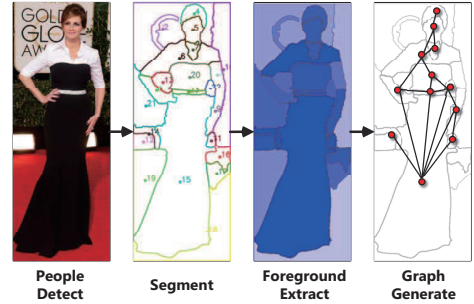}
\vspace{-0.05in}
\caption{Portrait graph representation.}
\label{fig:graph}
\end{center}
\vspace{-0.25in}
\end{figure}

We would like to take a typical photographing process as an example to
describe the functionality of each component and the system workflow.
Users create their personal portrait profiles using the \emph{Self Portrait Profile Generation} component
 and encode their portrait profiles to express their privacy requirements.
%Note that for privacy protection purpose, a user cannot use his/her portrait image directly as his/her portrait profile.
In our design, a set of vision feature vectors are extracted from subregions of a portrait image.
Both face and body features are extracted, in case that there may lack a clear front view of face.
We introduce a graph structure to encode the extracted vision features (feature vectors are properties of nodes) and
 use the graph as the portrait profile, as the examples in Fig.~\ref{fig:graph}.
Compared with uploading the original image,
 the feature graph shows a low risk on privacy leakage without lose of matching functionality,
 and are much more efficient for both computation and communication.
Besides, the graph representation is highly robust for pose changes of people and cameras.
%Different types of graph nodes own different types of feature vectors, \eg, face feature vector for Node No.5,  color and texture vectors for Node No.6.
%A \emph{label} is associated with each node to indicate its type.
%The selected featured should be invariant to scaling, rotation, and partially invariant to
%change in illumination and camera viewpoint. As the evolving of computer vision techniques, new features can conveniently be introduced to our system.
For each invisible user,
 his/her self portrait profile is generated once and for all until he/she updates it.
While the face features of a user remains the same, the user could change his/her outfits.
The portrait profile could be automatically updated when the user selfies
 or while he/she uses the phone with the frontal camera facing himself/herself.
In our advanced approach, transformed version of portrait graph are used to improve privacy protection.
With portrait graphs, we convert the people matching problem to graph matching problem.

Triggered by a photo shooting action,
 the \emph{Proximity Service} on cloud will be automatically notified and start to check
 if there are invisible people in the FOV of the photographer.
If any, the cloud will inform them and start the next step.
%Usually, cloud can search for the users that are located within certain distance threshold from the photographer.
Proximity service can be realized easily using common location service and onboard compass.
%It is not always convenient to recognize an accurate FOV, which needs parameters and the current orientation of the camera.
%In practice, the cloud could efficiently search invisible neighbors (located within a certain distance) of the photographer
% and take them as potential people appearing in the photo.
By restricting the number of potential matched invisible users,
 our system can achieve high matching accuracy and low overhead.

\begin{figure}[t!]
\begin{center}
\includegraphics[width=0.7\linewidth, clip,keepaspectratio]{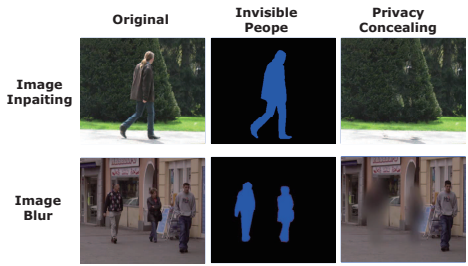}
\vspace{-0.05in}
\caption{Example of automatic privacy concealing (erase invisible people from photos) by image inpainting \cite{criminisi2004region} or blur.}
\label{fig:conceal}
\end{center}
\vspace{-0.25in}
\end{figure}

In the next step, after being informed by cloud,
 invisible neighbors upload their self portrait profiles to cloud
 (this could be done in advance to reduce the delay).
Meanwhile, the photographer detects all people in the photo and extracts their portrait profiles with the
\emph{FOV Portrait Profiles Generation} component,
 which works similarly to self portrait profile generation.
%Portrait profiles of multiple users may be extracted at one time.
These profiles will be uploaded to cloud as well.
%FOV profile generation can be completed by cloud when the cloud is trusted.
Then the \emph{Matching Service} will match portrait profiles of invisible users to portrait profiles from the photo
 and determine people that should be erased from the photo.
The graph matching algorithm will be discussed in detail in Section \ref{sec:matching}.
The matched results will be sent to photographer, and then the \emph{Privacy Concealing} component will erase the corresponding
 image regions of invisible people from the photo automatically by blurring
 or other more sophisticated techniques, like image inpainting \cite{komodakis2006image, criminisi2004region},
 to maximize the aesthetics.
We show an example of removing invisible people from photo in Fig.~\ref{fig:conceal}.
For the inpainting, we use the code from Criminisi's work \cite{criminisi2004region}.
After the removal, the photographer can store or share the photo using the cloud service.
Note that, based on the personal specification, the whole procedure works the same way for tagged users
 and the ''erase'' operation can be alternated to ''tag'' to augment many social applications.

In case there are dishonest photographers who don't complete the removal, \ourprotocol supports verification of removal.
All invisible users' portrait profiles have been uploaded to the cloud in the previous stage.
Once the photo is shared through Internet, the \emph{Verification Service} will check the photo as follows£º
the cloud first conducts a people detection on the photo and extracts all portrait profiles;
then the cloud matches these profiles with the cached profiles of invisible neighbors,
 if there is a matching, it can tell that the photographer didn't follow the protocol.
The verification process can be completed alone by the cloud without any interaction with users.
%Visual profile generation:
%1. Face profile. The most common face recognition algorithm is Eigenfaces introduced in \cite{turk1991eigenfaces}
%and Fisherfaces introduced in \cite{belhumeur1997eigenfaces}.
%Eigenfaces achieves approximately $96\%$ classification rate\cite{erkin2009privacy}.
%Both Eigenfaces and Fisherfaces transform face image into characteristic feature vectors in a vector space whose basis consists of Eigenfaces.
%The Eigenfaces are system parameters determined by a set of training images.
%Each face is represented as a vector in the face space by projecting the face image onto the subspace spanned by the  Eigenfaces.

%Recognition is conducted by searching the closest feature vector. The closeness is measured by Euclidean distance.

%\subsection{Main Idea}

%1. Visual distinguishability of people in a close proximity.
%2. The  order-independent of vector similarity

\subsection{Overview of Advanced System}
\vspace{-0.05in}

\begin{figure}[t!]
\begin{center}
\includegraphics[width=0.7\linewidth, clip,keepaspectratio]{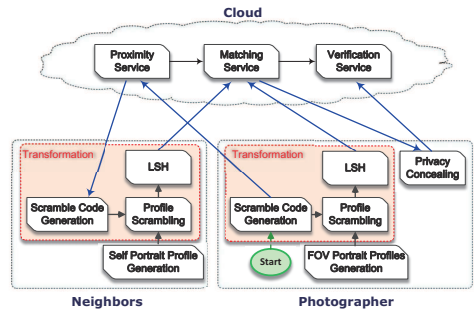}
\vspace{-0.05in}
\caption{Advanced system architecture.}
\label{fig:advance}
\end{center}
\vspace{-0.25in}
\end{figure}

The baseline protocol
%solves the problem that
%matching people in the photo with their privacy intentions and
%removing invisible people,
% thus
protects the visual privacy of people's portraits,
but exposes users' portrait profile (\ie feature vectors) to the cloud and even the eavesdroppers.
%However, users' portrait profile (\ie feature vectors)
% are exposed to the cloud and even the eavesdroppers.
With some feature vectors an adversary could have a chance to match them with existing photos or even reconstruct the photo.
In the advance approach, we retains the visual portrait privacy protection and
improve the system to protect users' portrait feature privacy and inference privacy (defined in Section \ref{sec:threat}).
The graph based profile matching scheme should be conducted in a privacy-preserving manner.
%Specifically, the portrait feature vectors of every invisible user
%(from both himself/herself and the photographer) should be protected from untrusted cloud and any other parties.
The core of the portrait profile matching algorithm is to measure the distance between vision feature vectors.
We cannot directly adopt existing privacy-preserving vector distance protocols based on homomorphic encryption \cite{katz2008predicate, sadeghi2010efficient} and gabled circuit\cite{sadeghi2010efficient} due to their large computation and communication cost.
In \ourprotocol we propose a highly efficient outsourced vector
 distance protocol (see Section \ref{sec:advanced}).
As shown in the red blocks in Fig.~\ref{fig:advance},
 combining a well designed scrambling scheme and locality sensitive hash,
 all invisible neighbors can secretly transform their vectors in a distance preserving way.
% and upload the portrait graph with transformed vectors to cloud, say the transformed portrait graph.
Then the cloud can measure vector distances using the transformed vectors
 and match transformed portrait graph with the same algorithm as in the baseline system.
Our scheme protects invisible users' portrait profiles (from both himself/herself and the photographer) with very little extra cost on the client side for generating transformed portrait graph.
But, it saves computation cost for the cloud, because the distance computation of high-dimensional real number vectors is converted to
distance of low-dimensional binary hash code.

The architecture of the advance system is presented in Figure~\ref{fig:advance},
 except vector transformation and verification,
 other components are the same as that in the baseline system.
Here, the verification is more challenging,
 because the cloud only knows the transformed portrait graphs of invisible users.
Without knowing the secret transformation,
 the cloud cannot compare them with portrait graphs directly extracted from the uploaded photo.
When an invisible user needs to check if his/her portrait has been removed,
 he/she need to start a verification and participate in the process as follows:
 the cloud sends all extracted feature vectors in a random order to the invisible user.
Note that, these feature vectors are supposed to belong to preserved visible people if the photographer is honest.
And the invisible user transforms them in the same way as his/her self feature vectors and sends the results transformed vectors to the cloud.
Then the cloud can compare preserved people in the photo and invisible neighbors using transformed portrait graphs,
 and detect the dishonest photographer.

%\subsection{Privacy Concealing and Verification}
%%In this section, we discuss the privacy concealing and verification strategy for both our basic and advanced design respectively.
%The privacy concealing is the same for both basic and advanced design.
%Once the cloud completes matching, it returns indices of invisible people in the photo to the photographer.
%The photographer device should automatically erase all invisible people by blurring
% or other more sophisticated techniques, like image inpainting \cite{criminisi2003object, komodakis2006image},
% to maximize the aesthetics.

\vspace{-0.05in}
\section{Portrait Profile Generation and Matching}
\vspace{-0.05in}
\label{sec:matching}
%\subsection{Matching Scheme}
%\label{sec:match}
%In \ourprotocol
% we determine whether a user has been photographed by matching his/her portrait profile with portrait profiles extracted from the photo,
% the matching score tells if two portrait profiles represent the same person.
%One of our main contribution is that we introduce a graph representation of portrait profile
% to achieve accurate, robust portrait matching.
%Moreover, an efficient algorithm is designed to measure the similarity of two portrait graphs (undirected graphs with node properties).
%In our system, we let the photographer do the profile graph generation while the cloud conduct the matching,
% because the portrait images of invisible users should not be exposed to the cloud or any other parties,
% but the costly matching work is preferred to be outsourced to the cloud.

\subsection{Portrait Profile Generation}
\vspace{-0.05in}
%For a system like \ourprotocol,
% a portrait profile should have the following properties:
% distinguishability in a local area,
% robustness for pose changes of people and camera ,
% well-formed structure for storage and matching,
% compatibility for the future development of vision feature description.
%We design a graph representation of portrait profile to have properties of distinguishability, robustness for pose change and compatibility for the future development of vision feature description.
After applying a people detection on a photo \cite{viola2004robust,leibe2005pedestrian},
 we obtain portrait images (including both people faces and bodies) from this photo as shown in the first subfigure of Fig.~\ref{fig:graph}.
Then portrait image can be segmented into adjacent regions by different colors and textures \cite{deng2001unsupervised}.
%There are a lot of work devoted to image segmentation and grouping in the field of computer vision, \eg, \cite{shi2000normalized}, \cite{deng2001unsupervised}, and \cite{arbelaez2011contour}.
%As a component of \ourprotocol,
% any unsupervised color and texture based segmentation solution can be utilized to obtain regions.
Given one portrait image,
 a graph $G=(V,E)$ can be constructed,
 where $V$ is a set of nodes representing segmented regions
 and $E$ are edges connecting any two regions that share a boundary.
%Fig.~\ref{fig:graph} and Fig.~\ref{fig:samples} illustrate some examples of our graph representations for portraits.
%\paragraph{Foreground extraction.}
%Now we have obtained the graph structure of a portrait image,
%However, some regions/nodes could be the background,
% which could deteriorate the matching correctness and efficiency.
Then we measure each node's confidence of being a part of the person and remove the node with low confidence to eliminate the background.
%The first cue is the extend that the region shares its boundary with the border of the portrait image.
% which gives an evidence that a region doesn't belong to the entity.
%The second cue is the distance from the mass center of a region to the center of the portrait image.
% which gives an evidence that a region belongs to the entity.
%The confidence of a node is obtained by fusing evidences using Dempster Shafer theory (DST) \cite{shafer1976mathematical},
% and we remove the node with low confidence to eliminate the background.
The confidence calculation is omitted due to space limitation.
% a reader can refer to \cite{longversion} for more detail.
Fig.~\ref{fig:graph} shows examples of foreground extraction and portrait graph generation,
 which provide more accurate graph representation of people portrait.
The result of foreground extraction can also be employed by the privacy concealing component
 as the accurate erase area to achieve better looking removal, as shown in Fig.~\ref{fig:conceal}.
%\paragraph{Node properties.}
%Segmentation not only yields the portrait graph structure,
%but also results in relatively consistent color and texture for each region.
For each region of portrait, vision feature vectors, \eg, face feature vector, color histogram and texture vector, are extracted as property of the corresponding node.
%For the node of human face, (\eg, Node No.5 in Fig.~\ref{fig:graph}),
% face feature vector, \eg, eigenfaces \cite{turk1991eigenfaces}, can be extracted.
%For body nodes, we can use feature vectors invariant to scaling, rotation and partially invariant to change in illumination and camera viewpoint,
% for example, color and texture.
%For each feature, various description vectors have been designed.
We will give more detail about node properties in the implementation section (Section.~\ref{sec:implementation}).

%\begin{figure}[t!]
%\begin{center}
%\includegraphics[width=0.88\linewidth, clip,keepaspectratio]{matrix.eps}
%\caption{The matrix recording possible matches of nodes in X and Y.}
%\label{fig:matrix}
%\end{center}
%\vspace{-0.15in}
%\end{figure}
\vspace{-0.05in}
\subsection{Portrait Graph Matching Scheme}
\vspace{-0.05in}

To achieve accurate and efficient portrait graph matching,
 there are several challenging issues should be addressed with low computation cost:
 graphs structure of the same person varies due to changing illumination condition and viewpoint;
% for example, the same skirt could be segmented to two nodes by a user or three nodes by a photographer;
 incomplete graphs could be produced due to occlusion;
 portrait profile could still contain some noisy nodes from background.
As a result, the matching algorithm should be elastic to node/edge division, aggregation, insertion and deletion,
 and robust to noise nodes.
%Although, there are many graph matching methods tailored for different applications \cite{wiskott1997face, isenor1986fingerprint, hu2013graph}.
Existing graph matching methods usually have application-oriented specifications \cite{wiskott1997face, isenor1986fingerprint, hu2013graph},
 \eg, assumptions about node numbers, graph structure and pre-knowledge of correspondences,
 make them difficult to be directly applied in this work.
To meet the critical requirements of portrait profile matching,
 we design a voting based strategy in which both the node similarity and graph structure are considered.

Let graph $G^x = (V^x, E^x)$ denote portrait profile $X$ (say produced
by a user) and $G^y = (V^y, E^y)$ denote portrait profile $Y$ (say
produced by a photographer).
Here $V^x = \{v^x_1, v^x_2,...,v^x_p\}$ and $V^y = \{v^y_1, v^y_2,...,v^y_q\}$.
Each node own some feature vectors as its property.
In order to improve matching accuracy as well as speed up the computation process,
 we add a \emph{label} for each node, which describes its type, for example, \emph{human face} or \emph{human body}.
Only nodes of the same type can be matched.
The distance between two nodes of different types is regarded as infinite.
%Thus the search space can be effectively pruned.
As human face is a strong feature to identify a person,
our matching scheme will firstly consider the matching between nodes labeled with "human face" (\eg, Node No.5 in Fig.~\ref{fig:graph}),
 then invoke an integrated graph matching.
%If we do not get clear face image, an integrated graph matching is invoked.
In this way, our method provides more accurate and robust matching than existing face recognition based methods.

\paragraph{Initialization.}
Let the similarity between nodes $v^x_i$ and $v^y_j$ be $\Sim(v^x_i, v^y_i)$,
 which can be obtained through measuring the distances between feature vectors of two graph nodes.
Note that, if $v^x_i$ and $v^y_i$ have different type labels, $\Sim(v^x_i, v^y_i)$ is set to zero.
In Section~\ref{sec:advanced}, we will discuss the details of privacy preserving vector distance computation.
%As the development of image processing techniques, more types can be introduced to improve the matching performance.
During the matching process, a matrix $M$ with $p$ rows and $q$ columns is built.
% as illustrated in Fig.~\ref{fig:matrix}.
Each entry $M_{ij} = \{f_{ij}, n_{ij}, c_{ij}\}$ of the matrix is a triple
 where $f_{ij}$ is a boolean flag indicating whether node $v^y_j$ is a possible match for node $v^x_i$,
 $n_{ij}$ caches the one-hop neighbor match information and $c_{ij}$ is a counter.
Details of these parameters will be presented in the following parts.
A match is represented as an assignment for all $\{f_{ij}\}$,
 where there is at most one $f_{ij}$ equaling $TRUE$ for every column $j$.
All $\{f_{ij}\}$ are initiated to $TRUE$.

%\begin{figure}[t!]
%\begin{center}
%\includegraphics[width=0.8\linewidth, clip,keepaspectratio]{match.eps}
%\caption{Example of portrait profile matching.}
%\label{fig:match}
%\end{center}
%\vspace{-0.15in}
%\end{figure}

After the initialization, our graph matching scheme consists of three stages.

\paragraph{Stage 1.} We eliminate wrong matches based on the similarities of node pairs.
If $\Sim(v^x_i, v^y_i)$ is above a pre-specified threshold $\xi_s$,
 the corresponding flag $f_{ij}$ is set to $TRUE$, otherwise, we eliminate this match by set $f_{ij}$ to $FALSE$.
%Also, if degree difference of the two nodes exceed a threshold $\theta_d$, the flag $f_{ij}$  is set to $FALSE$.
Note that, a node in $V^x$ does not necessarily have a possible match in $V^y$, thus there can be rows with all $FALSE$ flags.
After this stage, all node pairs with $TRUE$ flags are considered as \emph{candidate matches}.

\paragraph{Stage 2.} We explore the one-hop neighbor matching for each candidate match.
For each candidate match $(v^x_i, v^y_j)$, the neighbor sets of  them are denoted as $NE(v^x_i)$ and $NE(v^y_j)$.
%An example is shown in Fig.~\ref{fig:match}(a).
We find the most likely mapping from $NE(v^x_i)$ to $NE(v^y_j)$.
To achieve this, we firstly look for potential matches in matrix $M$ for each node in $NE(v^x_i)$.
%After Stage 1, there would still be multiple candidates for each node.
We then connect each node in $NE(v^x_i)$ with its matched nodes in $NE(v^y_j)$ with undirected edges.
% as shown in Fig.~\ref{fig:match}(b).
Nodes in both sets as well as the edges form a bipartite graph and the problem can be transformed to
 find a maximum match on the bipartite graph.
To address this problem, we apply the Hungary algorithm \cite{kuhn1955hungarian} which outputs a mapping from $NE(v^x_i)$ to $NE(v^y_j)$.
As mentioned above, the mapping is denoted as $n_{ij}$.
\begin{displaymath}
n_{ij}(v^x_a) = \left\{ \begin{array}{ll}
v^y_b & \textrm{if $v^x_a$ matches $v^y_b$}\\
\Phi & \textrm{if there is no match in $NE(v^y_j)$ for $v^x_a$}\\
\end{array} \right.
\end{displaymath}
where $v^x_a \in NE(v^x_i)$ and $v^y_b \in NE(v^y_j)$.

%\begin{displaymath}
%where $v^x_a$ \in $NE(v^x_i)$ and $v^y_b$ \in $NE(v^y_j)$
%\end{displaymath}

%We denote the subgraph induced by $v^x_i$ and its one-hop neighbors as $G^x_i$, the subgraph induced by $v^y_j$ and its one-hop neighbors as $G^y_j$.
%Then the similarity of the two neighbor sets are calculated as follows.

%If the similarity is above the threshold $\theta_s$, $(v^x_i, v^y_j)$ is retained to be candidate matches and otherwise we eliminate the mapping by setting flag $f_{ij}$ to $FALSE$.

\paragraph{Stage 3.}
We choose at most one assignment for each node in $V^x$ by a voting based scheme.
For each candidate match $(v^x_i, v^y_j)$,
 we build two trees rooted at $v^x_i$ and $v^y_j$ on graph $G_x$ and $G_y$ respectively.
 The two trees are traced in parallel on two graphs with the BFS method.
Here we restrict the tree growth to the constraint that,
once a node $v^x_k$ on $G_x$ and its matched node $v^y_g$ are appended to the trees,
the neighbors of $v^x_k$ which have not been included can be added to the tree only if they have matched nodes in $NE(v^y_g)$ according the recorded mapping $n_{kg}$.
%The insight of this scheme lies in that since elements in $\{n_{ij}\}$ have cached the matched subtrees with one-hop depth,
% in Stage 3 we attempt to assemble these subtrees to a bigger one.
% Due to the noises in different images and the viewing angle, portrait profile graphs for the same person have variations in their structures,
% thus the tree growth process may terminate without covering all nodes.
 When two trees have grown to the maximum size, we get a possible match for the subgraphs.
 In this approach, we propose a voting scheme to determine the best match.
 That is, for each candidate match $(v^x_k, v^y_g)$ on the two trees,
 we increase the counter value of $c_{kg}$ in entry $M_{kg}$.
 After trees of all candidate matches $(v^x_i, v^y_j)$ voted,
 we check the $c_{ij}$ in each entry $M_{ij}$ and retain the largest one for each column.
% Fig.~\ref{fig:match}(c) illustrates an matched example.
 Then matrix $M$ indicates a most likely match of $G^x$ and $G^y$ and the similarity between the two portrait profiles are calculated by integrating similarities of all matched nodes and edges.
{\small
\begin{eqnarray}
\Sim(G^x, G^y) = \frac{\sum_{f_{ij} = \text{TRUE}}\Sim(v^x_i, v^y_j)}{\parallel V^x \parallel + \parallel V^y \parallel} \nonumber
+ \frac{\sum_{e_{ab} \in E^x} \sum_{e_{cd} \in E^y}\delta(e_{ab}, e_{cd})}{\parallel E^x \parallel + \parallel E^y \parallel}
\end{eqnarray}
}
where
\begin{displaymath}
\delta(e_{ab}, e_{cd}) = \left\{ \begin{array}{ll}
1 & \textrm{if $f_{ac} = TRUE$ \& $f_{bd} = TRUE$}\\
0 & \textrm{otherwise}\\
\end{array} \right.
\end{displaymath}

\vspace{-0.05in}
\section{Improvement of Preserving Portrait Privacy}
\vspace{-0.05in}
\label{sec:advanced}

%In the basic system, since directly computing distance using original feature vectors
% will violate users' portrait feature privacy and inference privacy (defined in Section \ref{sec:threat}),
%In this section, we present our efficient outsourced privacy-preserving vector distance protocol,
% which is employed by the advance system.
%It hides users' vector values from untrusted server and other parties and prevents adversaries from
% measuring distance between private vectors and vectors from a dictionary.
%Only little extra cost is incurred than the basic approach with plain vectors.

%For practical system,
% the efficiency is as important as privacy.

% Existing work usually achieves outsourced privacy-preserving vector distance
% using homomorphic encryption \cite{katz2008predicate, erkin2009privacy} and gabled circuit\cite{sadeghi2010efficient}, which will cause unwanted high computation cost for users and cloud
% to encrypt high-dimension vision vectors and calculate distance homomorphically.
%Most of those solutions require multiple rounds of communication between the private input holders, \ie the photographer and neighbors.
%There is also a large communication cost for cipher blocks.
%\mynote{citation}
%The heavy overhead and long delay make those work not suitable for
% mobile applications.

%In this section, we propose a highly efficient outsourced vector distance protocol,
% which achieves the above privacy requirements while incurs little extra cost than the basic approach with plain vectors.

The main idea of our outsourced privacy-preserving vector distance protocol is to transform the original vectors to random vectors, meanwhile, preserve the distance among vectors.
Moreover, the transformation should be kept secret from adversaries.
In this way, the distance can be measured on transformed vectors as on original vectors (which means light-weight computation),
but the adversary cannot obtain the original vectors nor compute the distance between the transformed vectors and vectors from a dictionary
to infer the original ones.
%So the transformation should have the following properties:
% (1) is challenging.
Let the distance function of two vectors $\vx=(\vx_1,\vx_2,
\cdots),\vy=(\vy_1, \vy_2, \cdots) \in \vv$ be $\dis(\vx, \vy)$.
As shown in Fig.~\ref{fig:advance}, we design the transformation with two main building blocks: \emph{Profile Scrambling} and \emph{Locality Sensitive Hash (LSH)}.
The profile scrambling module works based on the observation that vector distances are \emph{dimension-order-independent},
 that is when we randomly change the dimension order of both $\vx$ and $\vy$ consistently to obtain scrambled $\vx'$ and $\vy'$,
 we have $\dis(\vx, \vy) \equiv \dis(\vx', \vy')$.
Once the scrambling order is kept secret, the original vectors are protected and a dictionary based inference is prevented.
In case there may be some dimension-dependent characteristics of vision feature vectors,
 \eg, in the color histogram the dimensions representing red component usually have large values,
 we employ the LSH module to transform the scrambled feature vectors into another low-dimensional vector space.
LSH hides the scrambled feature vectors from all parties and makes the statistic analysis on scrambled vectors infeasible,
 meanwhile it also preserves the distance among vectors.
Besides, lower-dimension vectors reduce the cost for vector distance computation and vector transmission.
On the other hand,
 changing the dimension order of $\vx$ randomly to $\vx'$ makes their hashes totally different,
 because there is a random distance between them.
Hence, the vector scrambling works like a random salt to strength the security property of LSH as well,
 that makes the dictionary attack against LSH infeasible.
Combining vector scrambling and LSH, we protect feature vectors of invisible users from untrusted cloud and other parties,
 and outsource most computation to the cloud in a secure and noninteractive manner.
%\begin{compactitem}
%\item Feature vectors of invisible users are well protected from untrusted cloud and other parties.
%\item Compared to homomorphic encryption or garble circuit based methods, our solution is  much more efficient.
%\item Most computation cost are outsourced to the  cloud and no interaction is required during the matching process.
%\end{compactitem}

In the following subsections,
%we present the detail of our privacy-preserving vector distance protocol.
we will introduce the LSH based vector distance measurement as a preliminary,
then present our privacy-preserving vector distance protocol.

\vspace{-0.05in}
\subsection{LSH based Vector Distance Measurement}
\vspace{-0.05in}

% in this work we present a new vector distance measurement scheme based on Locality Sensitive Hashing (LSH), achieving much better privacy while retaining low overhead.
%
%LSH has already been applied in solving problems like the approximate nearest neighbor search \cite{datar2004locality}.
The key insight behind LSH is that it is possible to construct
 hash functions such that close vectors will have the same hash value with higher probability than vectors that are far apart.
% Given a metric space ($\vv, \dis$), a family $\hf=\{h:\vv \to U \}$ is ($r,cr,p_1,p_2$)-sensitive if for any $\vx,\vy \in \vv$ the following two conditions hold:
%\begin{itemize}
%\item If $\dis(\vx,\vy) \leq r$, then $Pr_{h\in\hf}[h(\vx)=h(\vy)] \geq p_1$,
%\item If $\dis(\vx,\vy) \geq cr$, then $Pr_{h\in\hf}[h(\vx)=h(\vy)] \leq p_2$.
%\end{itemize}
Different LSH functions are designed for various distance metrics, \eg, Euclidean distance, Hamming distance, cosine distance.
In our system, we use the commonly used Euclidean distance
$\dis_E(\vx, \vy) = \sqrt{\sum_i(\vx_i-\vy_i)^2}$.
Particularly, for high-dimensional vector space $\vv = R^D$ with Euclidean distance, an LSH function is defined as follows \cite{datar2004locality}:
\begin{align}
\lsh(\vx) = <h_1(\vx), h_2(\vx),\cdots,h_m(\vx)>\\
h_i(\vx) = \left\{ \begin{array}{ll}
1 & \textrm{if  $\frac{\va\cdot \vx}{W} \geq 1, i=1,2,\cdots,m$}\\
0 & \textrm{otherwise}\\
\end{array} \right.
\end{align}
where $\va\in R^D$ is a random vector with each dimension chosen
independently from the standard Gaussian distribution $N(0,1)$.
Here each $h_i$ is an atomic LSH function, and the LSH function $\lsh$
generates a hash vector of the input vector by concatenating $m$
scalar atomic hash values. The window size $W$ and $m$
control the distance range that the mapping is sensitive to.
In the advance system,
 the cloud determines the hashing function $\lsh$ and publishes it to all participants.

According to the definition of LSH,
 the differences between hash vectors indicate the distance between original vectors.
%As our graph matching scheme is not sensitive to the concrete distance value
% but only the relative distance or similarity between feature vectors,
In this work we apply the Hamming distance between hash vectors to approximate the distance between original vectors.
Our experimental results show that the Hamming distance between hash vectors is nearly monotonic to distance measurement between original vectors.

%With LSH, the lower-dimension hash vectors protect the true values of input feature vectors
% and also enable highly efficient matching for the cloud.
% \mynote{add analysis here.}
%But there still exists a risk of privacy leakage, since the hash is deterministic.
%For example, an adversary can connect the same invisible person appeared in different photos by
%comparing the hash vectors.
%Or the adversary could even use a photo collection to generate a dictionary of hash vectors
%and guess the invisible portrait content by comparing the hash vectors.
%Each participant communicating with the cloud through a secure channel can prevent adversaries outside the system from accessing hash vectors,
%but it cannot prevent untrusted cloud.
%To resist the hash vector comparison based attacks,
% we further improve \ourprotocol by avoiding deterministic LSH outputs.

\vspace{-0.05in}
\subsection{Outsourced Privacy-preserving Distance Computing}
\vspace{-0.05in}

%As illustrated in Fig.~\ref{fig:advance}, to outsource portrait matching to could in the privacy-preserving manner,
% each participant should transform his/her vectors by first scrambling the their dimension order and then conducting LSH on
%  the scrambled vectors.
%Portrait graph with transformed vectors are uploaded to cloud for portrait matching.
Here, we assume that each user share a secure communication channel with the cloud.
Then the transformed vectors are protected from other participants.
In a specific round of photographing, to preserve distance between transformed vectors,
 the challenge is that all participants (photographer and invisible neighbors)
 must scramble their feature vectors in a consistent order individually and secretly.
We refer to the scramble order as scramble code $\code$.
To achieve the same $\code$, all participants first need to generate a same random seed $R$ secretly.
The multi-user agreement protocol requires that the untrusted cloud cannot learn the random seed and the scramble code,
 although it controls all communications between users.
%Dynamic parties: it is better if any party can join and leave at any given time without much extra work for other parties.

\paragraph{Random number exchange.}
There are many well-designed group key agreement protocols \cite{burmester1995secure, lee2006distributed}, 
 but most of them require multiple communication rounds among participants,
 which could cause long delay.
Utilizing the honest-but-curious cloud and secure communication channels between the cloud and users,
 we adapt the practical distributed group key agreement protocol proposed in \cite{burmester1995secure} to
 achieve round optimum and computation efficient random number agreement.
Let $U_1, \cdots, U_n$ be a dynamic subset of all users who want to generate a common random number
 and our protocol is presented in Algorithm~\ref{algorithm:random}.
With this protocol, the photographer and his/her invisible neighbors can obtain the same random number,
 while the cloud learns nothing about the random number.
%There are some distributed round-efficient group key exchange protocols, \eg, \cite{burmester1995secure, tzeng2000round}.
%\cite{tzeng2000round} achieves round optimum and is secure against both passive and active adversaries under the random
%oracle model. It also releases no useful information to passive adversaries
%and achieves fault tolerance against any coalition of malicious participants.

\begin{algorithm}[h]
\renewcommand{\algorithmicrequire}{\textbf{System Initialization:}}
\renewcommand\algorithmicensure {\textbf{Runtime:} }
\caption{Random Number Agreement.}
\label{algorithm:random}

\begin{algorithmic}[1]
{\small
\REQUIRE ~~\\
Cloud generates and publishes system parameters:
1) a large prime number $p=\Theta(2^{cN})$, a constant $c \geq 1$,
 $q=\Theta(2^N)$ and $g \in Z_p$ of order $q=\Theta(2^N)$.\\
Each user $U_i$ generates his private parameter $a_i \in Z_q$ and public parameter $b_i=g^{a_i} \mod p$ and sends $b_i$ to the cloud.\\
Cloud checks that $b_i^q \equiv 1 \mod p$ for all $i=1,\cdots,n$.

\ENSURE ~~\\
%Random number $R$.

\STATE Cloud arranges $n$ users' indices in a cycle and sends $b_{i-1}$ and $b_{i+1}$ to each $U_i, i=1,\cdots,n$.

\STATE Each $U_i, i=1,\cdots,n$ computes $c_i$ and sends it to cloud \label{step}
\begin{equation}
    c_i = (b_{i+1}/b_{i-1})^{a_i} \mod p.
\end{equation}

\STATE Cloud sends $c_1, \cdots, c_n$ to each $U_i, i=1,\cdots,n$.

\STATE Each $U_i, i=1,\cdots,n$ computes the random number
\begin{equation}
    R_i = (b_{i-1})^{na_i}\cdot c_i^{n-1}\cdot c_{i+1}^{n-2}\cdots c_{i-2} \mod p.
\end{equation}
Although each user generates $R_i$ individually, all $R_i$ equal to the same random number
\begin{equation}
    R = g^{a_1a_2 + a_2a_3 +\cdots +a_na_1} \mod p.
\end{equation}
}
\end{algorithmic}
\vspace{-0.05in}
\end{algorithm}
\vspace{-0.05in}

\paragraph{Scramble code generation.}
After obtaining consistent random seed $R$,
 each participant generates the scramble code using Algorithm~\ref{algorithm:code},
 and rearranges the dimension order of each feature vector according to the scramble code.

\begin{algorithm}[h]
\renewcommand{\algorithmicrequire}{\textbf{Input:}}
\renewcommand\algorithmicensure {\textbf{Output:} }
\caption{Scramble code generation.}
\label{algorithm:code}

\begin{algorithmic}[1]
{\small
\REQUIRE ~~
Vector dimension $N$; Random number $R$;
Sorted set $S=\{1,2,\cdots,N\}$;

\ENSURE ~~
Scrambled dimension sequence $\code$;\\

\FOR{$k=N-1$; $k>=0$; $k--$}
\STATE $i = R/k!$;\\
\STATE $\code[N-k]=S[i]$;\\
\STATE Remove $S[i]$ from $S$;\\
\STATE $R = R \mod k!$;\\
\ENDFOR
\RETURN $\code$;
}
\end{algorithmic}
\vspace{-0.05in}
\end{algorithm}

As illustrated in Fig.~\ref{fig:advance}, after scrambling feature vectors,
 the photographer and invisible neighbors apply LSH to scrambled vectors to get transformed vectors for current round.
The cloud can simply use the transformed vectors to compute distance and conduct the same graph matching algorithm
 as in the basic scheme.
While the membership doesn't change, the random number remains the same.
In this case, the photographer can use the same random number to generate transformed vectors for new photos,
 and all invisible neighbors do not need any recalculation.
When the membership changes, the cloud can insert/remove users into/from the exiting ring of Algorithm~\ref{algorithm:random}
 to update the random number for a new round.
Note that, in this case, most users do not need to recalculate the Step~\ref{step} in Algorithm~\ref{algorithm:random}.
Based on our evaluation, the runtime for random generation time is usually only 0.014s, which is negligible for human movement.
Once the random number is updated, the system achieves randomized transformation outputs for the same feature vector in different rounds.
%Also, we tailor it to reduce the cost for dynamic membership.

%In summary \ourprotocol enables removing people who are unwilling to be
% photographed from captured photos in a privacy preserving manner.
%We designed two variants of the protocol, basic and advanced.
%In the first variant, \basic (Figure~\ref{fig:basic}),
%both photographer and his/her neighbors upload portrait feature graph
%explicitly.
%Portrait features of each invisible user are uploaded once and for all.
%While the face features of a user remains the same, the user could change his/her outfits.
% The portrait profile could be automatically updated when the user selfies
% or while he/she uses the phone with the frontal camera facing himself/herself.
%All overhead while photographing is outsourced to the cloud.

%The second variant,\advanced (Figure~\ref{fig:advance}),
% retains the visual portrait privacy protection,
% and are improved to provide portrait feature privacy and inference
% privacy protection.
%It costs clients little extra overhead for generating encoded feature
%vectors but saves computation cost for the cloud, because the distance
%computation of high-dimensional real number vectors is converted to
%hamming distance of low-dimensional binary hash code.

\vspace{-0.05in}
\section{Prototype Implementation and Evaluation}
\vspace{-0.08in}
\label{sec:implementation}

%We implement prototype systems of both variants and compare them with each other and other related solutions for comparison.
%answer the following questions:
%(1) what is the accuracy of the graph-based portrait matching? will the encoding mechanism reduce the matching accuracy?
%(2) what is the cost for the photographer and neighbors in the basic mechanism for visual privacy? and what is the cost for cloud to provide the service?
%(3) what is the benefit and extra overhead for the advanced mechanism?
%(4) what is the cost of other alternative schemes that could be used to implement the privacy-preserving profile matching?

\subsection{Prototype Implementation}
\vspace{-0.05in}
%To answer these questions, we implement both variants and some
%alternative methods to evaluate each component of our design.
%The image processing components (face/pedestrian detection, portrait
%profile generation) are shared by both variants.
%These components are realized with C++ to achieve better computation
%efficiency.
We implement prototype systems of both variants.
To support automatic people detection, we implement the most popular face detection \cite{viola2004robust}
 and pedestrian detection \cite{leibe2005pedestrian} algorithm based on the library of OpenCV.
%\cite{OpenCV}.
To generate portrait graph,
%we need image segmentation to get the graph structure and
% feature extraction to get the node properties.
%Many segmentation methods exist in the computer vision field
%\cite{deng2001unsupervised, arbelaez2011contour}.
%In our prototype,
we adopt JSeg \cite{deng2001unsupervised} for image segmentation.
% because it achieves good results with acceptable computation cost.
%The color threshold of JSeg is set to 100 and the merge threshold is set to 0.6, which work fine in all evaluations.
%There are many sophisticated feature descriptors with large sizes and may incur high computation cost.
%For the system efficiency,
% we use light weight feature vectors as the node property with little loss of accuracy.
Leveraging the library of MPEG-7,
%\cite{MPEG},
 a 48-byte eigenfaces vector\cite{turk1991eigenfaces} is extracted as the property of a node labeled with face,
 and a 64-byte color histogram vector and a 20-byte texture vector (edge histogram with 4 blocks and 5 orientations)
 are extracted as the property for other nodes.
%Specifically, the color vector is the joint color histogram for the H and S channels with $8\times8$ bins;
%the texture vector is an edge histogram with 4 blocks and 5 orientations.
%Although our prototype uses these features,
\ourprotocol is compatible with any other vector-based feature descriptors.
Obtaining the matching results, invisible people are removed from the photo by blurring and inpainting \cite{criminisi2004region},
 as shown in Fig.~\ref{fig:conceal}.
Except the image processing,
 all other building blocks are realized using Java, including portrait graph matching,
 LSH, random number agreement, vector scrambling and the messaging module.
The client side are developed as an app on Android system for case study.
A user starts this app by inputting his/her portrait profile via selfieing.
% and choose his/her status from ''invisible me'', ''tag me'' or ''do nothing''.

%We implement the messaging module based on Wi-Fi communication,
% \kbnote{both AP and AdHoc modes are supported.}

\vspace{-0.05in}
\subsection{Case Study and Experiment setup}
\label{sec:case}
\vspace{-0.05in}
%\ourprotocol is able to match surrounding people with those in the photo.
%For invisible users, the matching can be conducted in a
%privacy-preserving way to remove them from photos,
%but for visible users, our techniques can also be used to tag them automatically.
%While the former function protects people's portrait privacy,
% the latter function could be fun and useful in many social scenarios
% (like parties and conference)
% and give  photographers incentive to participate in this system.
%We build an app of \ourprotocol which realizes both functions.
%The client side are developed as an app on Android system for case study.
%Our app is able to match surrounding people with those in the photo,
% and remove invisible people or label tagged people automatically.
%A user starts this app by inputting his/her portrait profile via selfieing his/her current look,
% and choose his/her status from ''invisible me'', ''tag me'' or ''do nothing''.
%A photographer can use this app to capture photos, which processes images according to the matching results.

%\begin{figure}[h]
%\begin{center}
%\includegraphics[width=0.8\linewidth, clip,keepaspectratio]{samples.eps}
%\caption{Sample portrait images and their portrait profile generation from experiment, including the original image, detected foreground and portrait profile graph. (The faces are blurred for the purpose of anonymity.)}
%\label{fig:samples}
%\end{center}
%\vspace{-0.15in}
%\end{figure}

To test the practicality and efficiency of \ourprotocol
 the evaluation is conducted in a crowded real-life scenario:
 a networking workshop with more than 50 attendees in a $200m^2$ meeting hall.
%As same as the case in real life, there are many people who are not \ourprotocol users.
10 volunteers (4 female and 6 male) acted as invisible users and also photographers.
Within one day, the volunteers took photos freely and our system recorded the cost and photos.
After the experiment, we got 208 photos.
1326 pedestrians are detected which belong to 42 individuals (7
 female and 35 male), but only 412 faces are detected.
%The reason is that, pedestrian detection is much more robust from different view points,
% but face detection requires the frontal face of people.
It implies that
%pedestrian detection is much more robust from different view points£¬
% methods only using face recognition to remove invisible users
% are not sufficient for portrait privacy protection,
a whole body detection and description (\eg, our graph model) is necessary.
%Fig.~\ref{fig:samples} shows some sample pedestrian images and their portrait graphs extracted by our system automatically.
We manually labeled all captured people as the ground truth for the following evaluations.
%In our evaluations, we do not consider those people in photo who cannot be detected by our prototype,
% since they are usually very small or occluded badly.
%Besides, the detection rate could be improved with more sophisticated people detection algorithm,
% which is out of the scope of this work.

\paragraph{Experiment setting.}
In the experiments,
 we use three types of phones as clients: HTC G10 (1024Hz CPU and 768M RAM),
 HTC G23 (1536Hz CPU and 1G RAM) and HTC New One (1741Hz CPU and 2G RAM).
One laptop is used as the cloud: ThinkPad X1 with i7 2.7GHz CPU and 4GB
RAM.
%In our evaluation, HTC New One is about two times faster than HTC G10,
% and HTC G23 is a little slower that HTC New One.
%For simplicity, in the following evaluation, we present only the result from HTC G23.
%All similarities are normalized to the range $[0,1]$.
Based on our extensive evaluation, to achieve the tradeoff between matching efficiency and accuracy,
 we set the parameter of the matching methods as $\xi_s=0.5$ and the parameters of LSH as $W=3$ and $m=128$ (the hashed vector is 128 bit).
%Before experiments start, we need to decide the parameter $\xi_s$ of the matching methods,
% which is the threshold to eliminate unmatched nodes in Stage 1 (in Section \ref{sec:matching}).
%We conduct pair-wise portrait graph matchings on the dataset with different settings of $\xi_s$.
%When $\xi_s$ increases from 0 to 0.5, average matching time for a pair of portrait graphs is cut down by $30\%$,
% with little hurt to the similarity measurement.
%But while $\xi_s$ continues to increase,
% it eliminates more and more true match pairs.
%As a result,
% we set $\xi_s=0.5$ in the rest of our experiments.
%For better accuracy of distance measurement,
% we set the parameters of LSH as $W=3$ and $m=128$, and the hashed vector is 128 bit.
%We also need to determine the parameter of the LSH algorithm.
%The accuracy of distance measurement using LSH increases with bigger $m$ (longer hash code) and smaller $W$ (smaller window).
%According to the analysis in \cite{datar2004locality} and the statistics of our dataset,
% we set the $W$ to 3 and $m$ to 128, then the hashed vector is 128 bit.
%\mynote{add figure for the parameters.}
For the random number generation, $N$ is set to 512, which provides sufficient protection for the random number agreement protocol.
The analysis of parameter setting is omitted due to space limitation.
% and a reader can refer to \cite{longversion} for detail.
The the following evaluation,
 we denote the implementation of the baseline system as \basic
 and the implementation of the advance system as \advanced.
\vspace{-0.05in}
\subsection{Matching Accuracy}
\vspace{-0.05in}
Here we investigate the most important metric,
the portrait matching accuracy of both variants, which determines the correctness of invisible people removal.
%and visible people tagging.

\begin{figure*}[ht]
\begin{minipage}[t]{0.28\linewidth}
\centering
\includegraphics[width=0.9\linewidth, clip,keepaspectratio]{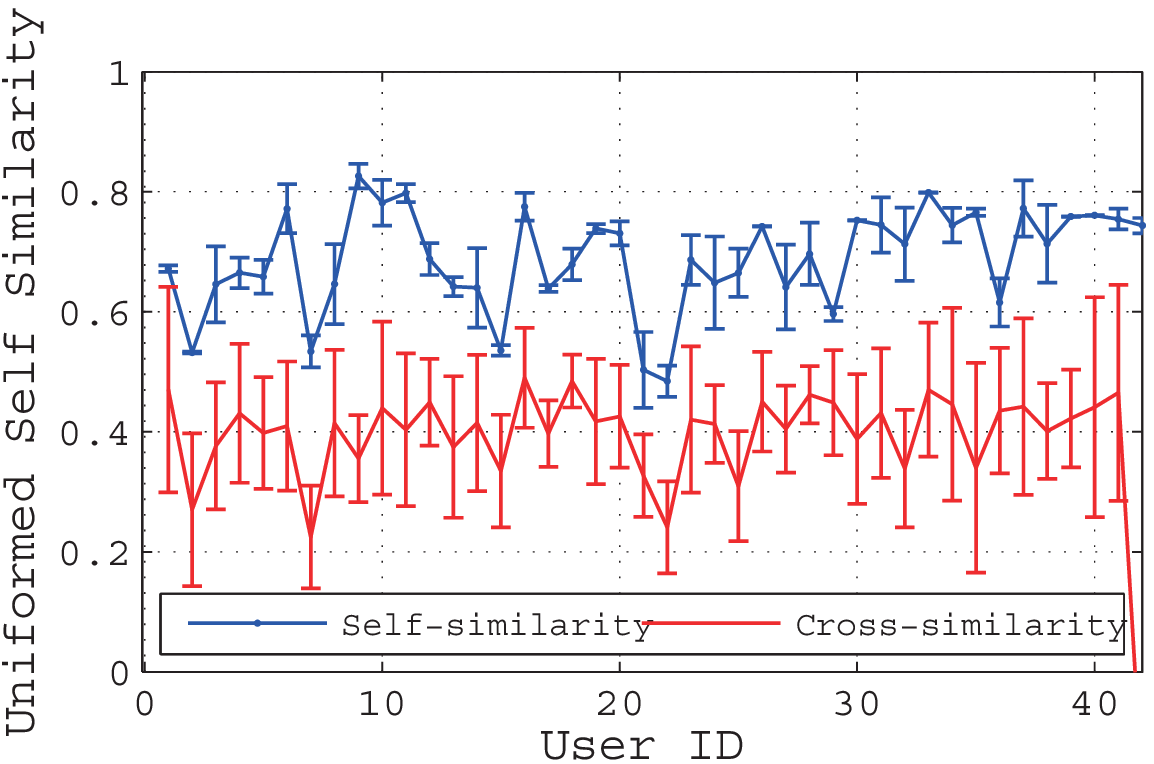}
\caption{Portrait similarity variances.}
\label{fig:self-sim}
\end{minipage}
\vspace{-0.1in}
\hfill
\begin{minipage}[t]{0.28\linewidth}
\centering
\includegraphics[width=0.9\linewidth, clip,keepaspectratio]{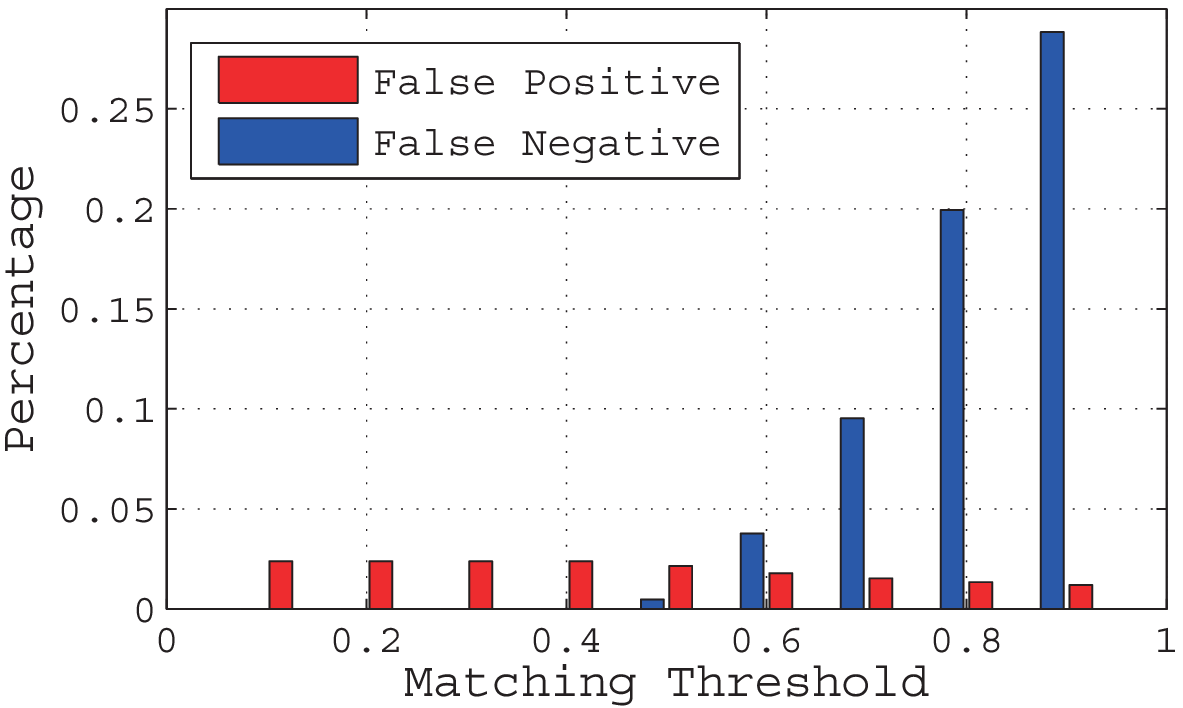}
\caption{FP and FN  in basic scheme}
\label{fig:false-basic}
\end{minipage}
\vspace{-0.1in}
\hfill
\begin{minipage}[t]{0.28\linewidth}
\centering
\includegraphics[width=0.9\linewidth, clip,keepaspectratio]{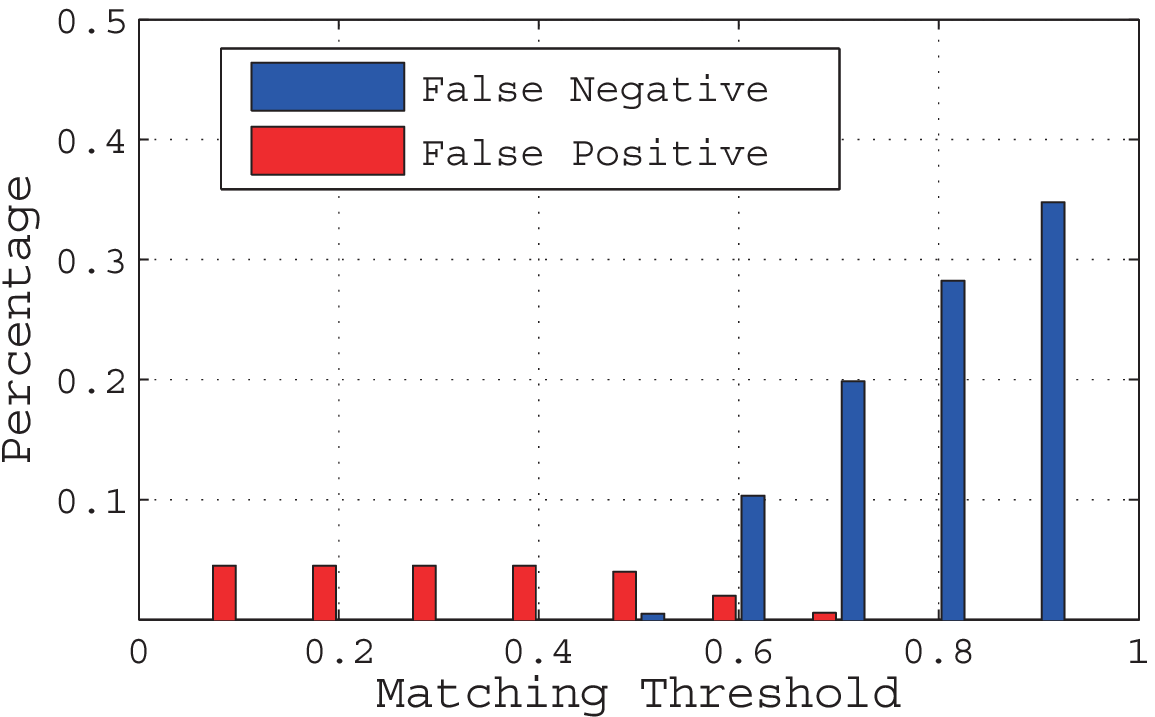}
\caption{FP and FN in advanced scheme}
\label{fig:false-advance}
\end{minipage}
\vspace{-0.1in}
\end{figure*}

%\begin{figure}[h]
%\begin{center}
%\includegraphics[width=0.6\linewidth, clip,keepaspectratio]{people_self_sim.eps}
%
%%\subfigure[V-Ads data set.]{ \label{fig:self-sim-social}
%%\includegraphics[width=0.6\linewidth, clip,keepaspectratio]{logo_self_sim.eps}}
%\caption{Portrait similarity variances.}
%\label{fig:self-sim}
%\end{center}
%\end{figure}

We start by examining the consistency and distinguishability of user's
portrait graph
 by self-similarity (similarity between the same entity's portrait graphs)
 and cross-similarity (similarity between different entities' portrait graphs).
In this evaluation, we remove the face property since it is highly distinctive
 but cannot always be obtained.
Figure.~\ref{fig:self-sim} presents the evaluation results using the dataset.
The upper blue line stands for mean self-similarity for each entity,
 and the lower red line is mean cross-similarity between this entity
 and all other entities.
We notice that, generally portrait graph has a good consistency, \ie,
high self-similarity and small variance.
And the obvious gap between self-similarity and cross-similarity
 shows a good distinguishability.
In fact, in most cases, portrait graph can provide accurate matching
 without face features,  which implies better privacy protection.

Then, we analyze the matching correctness by analyzing all possible combinations using the dataset.
A false negative (FN) happens when a user A is invisible, but not removed from the photo
due to a match score lower than a match threshold $\theta_s$.
A false positive (FP) happens when user A is invisible, but another visible user C is removed due to
 a higher match scores than both the threshold and A's score.
In \basic matching is conducted on portrait graph with plain feature vectors.
Fig.~\ref{fig:false-basic} illustrates the  percentage of
FN and FP changing with different threshold $\theta_s$.
By selecting the threshold $\theta_s=0.5$,
 \basic achieves $0.5\%$ false negative and $2.1\%$ false
 positive without using any face property.
With face property, the false negative decreases to about $0.1\%$ and
false positive is less than $1\%$.
In \advanced feature vectors are transformed by scrambling and LSH.
While the scrambling retains the accurate distance between vectors,
LSH could cause some accuracy loss.
Will the transformation reduce the matching accuracy?
With appropriate parameters $m=128$ and $W=3$,
\advanced achieves comparable accuracy with \basic, as shown in
Fig.~\ref{fig:false-advance}.
When  $\theta_s=0.5$,
 the false negative is about $0.7\%$ and the false positive is about
 $2.9\%$ without any face property.
So, both variants support accurate matching and our vector transformation achieves good portrait feature privacy
 protection with little accuracy loss.
%\textbf{Summary:}
% \ourprotocol (both the basic and advanced) support accurate matching,
% which is capable for portrait privacy protection and many other
% tagging applications.
%Our vector encoding method achieves good portrait feature privacy
% protection with little accuracy loss.
Based on the evaluation, in the rest experiments,
 the threshold $\theta_s$ is set to 0.5.

%\begin{figure}[h]
%\begin{center}
%\includegraphics[width=0.6\linewidth, clip,keepaspectratio]{false_p.eps}
%
%\caption{Probability of false positive and false negative in \basic without face property.}
%\label{fig:false-basic}
%\end{center}
%\end{figure}

%\begin{figure}[h]
%\begin{center}
%\includegraphics[width=0.6\linewidth, clip,keepaspectratio]{false_p.eps}
%
%\caption{Probability of false positive and false negative in \advanced without face property.}
%\label{fig:false-advance}
%\end{center}
%\end{figure}

\vspace{-0.1in}
\subsection{Micro Benchmark}
\vspace{-0.1in}

%We evaluate the cost of \ourprotocol by measuring the communication
%and computation cost of each component.

\paragraph{Communication cost.} In \basic each face node takes 48B and each other node takes 84B.
The size of the portrait graph depends on the node number $k$.
For most applications, $k \leq 10$ is sufficient,
so the communication cost for each portrait is 0.82KB.
In \advanced after encoding, each vector is hashed to 128 bits,
 which reduces the size of a portrait to 0.15KB.
Protocol \advanced requires extra communication for random number agreement,
 which is only about 0.19KB.
\ourprotocol
costs each participant less than 1KB data transmission to
enable portrait privacy protection.
The cost for a photographer depends on the people number in the captured photo,
 but in most cases (with less than 10 people in photo), less than 10KB overhead is incurred, which is much less than a photo.
In general, \ourprotocol achieves much smaller transmitted data size and better privacy protection
 than transmitting the image itself.

%\begin{figure}[ht]
%\begin{center}
%
%\subfigure[V-Social data set.]{ \label{fig:self-sim-social}
%\includegraphics[width=0.6\linewidth, clip,keepaspectratio]{people_self_sim.eps}}
%
%%\subfigure[V-Ads data set.]{ \label{fig:self-sim-social}
%%\includegraphics[width=0.6\linewidth, clip,keepaspectratio]{logo_self_sim.eps}}
%
%\caption{Portrait similarity variances.}
%\label{fig:self-sim}
%\end{center}
%\end{figure}

\paragraph{Computation cost.}
In \basic the computation cost is composed of portrait graph
generation on the client side, and portrait matching on the cloud.
\advanced costs extra computation for random number agreement and vector transformation by scrambling and LSH.
The runtime is only about 3 ms to transform ten 64-dimension feature vectors.
Table.~\ref{table:time} presents all the decomposed computing time.
It shows that,
 the major computation delay is caused by image processing.
%Portrait graph generation includes portrait detection, segmentation and feature extraction.
For a participant, it only needs to be executed once for the profile setup;
for a photographer, it needs to be executed for every captured photo.
The runtime of portrait detection and segmentation depend on the
resolution and complexity of the photo,
but the detection and segmentation results are not sensitive to scaling.
Hence, in our prototype all images are scaled to about 240,000 pixels.
%We run portrait graph generation on three types of phones.
For the photographer, on average it takes about 0.4s to conduct face
and pedestrian detection.
Given a portrait image/subimage,
% as plotted in Fig.~\ref{fig:runtime-gen},
 the processing time of segmentation and feature extraction is about 2.6s.
%increases with the image complexity, \ie region number after segmentation.
On average, there are 28.2 regions of each portrait.
% it takes about 2.6s to process one 240,000 pixel image.

\begin{table}[hp]
\vspace{-0.08in}
\caption{Microbenchmarks of Runtime (in second)}
\label{table:time}
\centering
{\footnotesize
\begin{tabular}{|l|c|c|c|}
%\hline
%\multicolumn{4}{|c|}{HTC G10}\\
%\hline
%& Min & Mean & Max\\
%\hline
%Segmentation & $1.1$ & $4.2$  & $10.5$\\
%\hline
%Extraction & $0.09$ & $0.5$ & $2.8$\\
%\hline
%Random-Gen & 0.013 & 0.018 & 0.021 \\
\hline
\multicolumn{4}{|c|}{Client}\\
\hline
Segmentation & $0.5$ & $2.4$ & $8.1$\\
\hline
Extraction & $0.02$ & $0.25$ & $1.3$\\
\hline
Random-Gen & 0.012 & 0.014 & 0.017 \\
\hline
%\multicolumn{4}{|c|}{HTC New One}\\
%\hline
%Segmentation & $0.5$ & $1.9$  & $6.0$ \\
%\hline
%Extraction & $0.05$ & $0.21$ & $0.68$\\
%\hline
%Random-Gen & 0.013 & 0.013 & 0.014 \\
%\hline
\multicolumn{4}{|c|}{Cloud}\\
\hline
Matching (basic) & $0.015$ & $0.037$ & $ 0.079$\\
\hline
Matching (advanced) & $0.006$ & $0.01$ & $ 0.039$\\
\hline
Random-Init &  1.31 & 0.9 & 1.57 \\
\hline
\end{tabular}}
\vspace{-0.1in}
\end{table}%%%%%%%%%%%%%%55 end of small

Compared with the image processing, the runtime of graph generation and matching is nearly negligible.
On the client side, only extra 0.014s runtime is required for random number generation in
 \advanced. On the cloud side, the time needed to match a pair of portrait graph
 is only about 0.04s in \basic and decreases to 0.01s in \advanced due to the hashed feature vector.
%To initialize the random number generation,
The cloud also needs 0.9s to generate system parameters for random number generation.
The millisecond-level portrait graph transmission delay is negligible too.
So the total computation delays for both variants are about 3s on the client
 and 1s on the cloud, which results a 4s system computation delay.
%depends on the network type (\eg, Wi-Fi, 3G) and condition.
%No mater what transmission techniques are applied,
% transmitting a portrait graph of size 1KB will only incur a millisecond-level delay.

Now we've learned the magnitude of the time cost for each component,
 the overall delay also depends on the number of co-located invisible neighbors.
With more active peers sending privacy requests, the matching cost will increase,
 but compared to the image processing cost, the matching cost on the cloud side is still quite small.
Besides, the power consumption caused by our protocol (second-level computation) is much smaller than that caused by photo capturing itself.

\vspace{-0.05in}
\subsection{Case Evaluation}
\vspace{-0.05in}
We conduct the case based evaluation as
described in Section \ref{sec:case}.
%There are 10 invisible users in a local area.
%The 208 photos captured them for 375 times in total.
When there are invisible users in the photo,
 the false negative rate was about $1.4\%$ and the false positive rate was $0.9\%$.
But when there are no invisible users in the photo,
 the false positive rate raises to $4.9\%$ due to the absent of any true match users,
 and the threshold 0.5 was not high enough to exclude all false match.
%with face features [TBD]
And the average time for successful invisible people removal is about 4 seconds.

\vspace{-0.05in}
\subsection{Compare with Alternative Solution}
\vspace{-0.05in}
%We propose an outsourced privacy preserving distance computation method
% using vector scrambling and LSH.
For comparison purpose, we also realize private Euclidean distance computation
 using a partial homomorphic encryption (Paillier encryption)
 in the SMC manner (e.g. the method used in \cite{katz2008predicate}).
Using the same computer and test images, the Paillier-based method takes about 0.5s for feature vector encryption
 and 1.8s for portrait matching between a pair of feature vectors.
But with our approach, the transformation cost is negligible and the matching cost is only 0.01s.
Besides, our method requires no interaction during the matching process.
The comparison shows the a significant efficiency of our system.

%\section{Case Study and Evaluation}
%\label{sec:application}
%\input{app.tex}

%\section{System Refinement}
%\label{sec:refine}
%\input{refine.tex}

\vspace{-0.16in}
\section{Related Work}
\vspace{-0.08in}
\label{sec:related}
\paragraph{Visual Privacy Protection}
There is a trivial solution to protect image content.
Blacking out private contents, \eg human faces,
  thwarts any possible violation of owners' privacy.
For example, systems like Blinkering Surveillance \cite{senior2005blinkering}
 and \cite{zhang2005hiding} use computer vision methods to hide sensitive contents from video frame.
% conceals persons in circumstantial video image.
%\cite{newton2005perserving} introduces an algorithm to protect the face privacy
% of individuals in video by blurring faces.
%Blinkering Surveillance \cite{senior2005blinkering} uses computer vision methods
% to hide superfluous details.
%particularly identity in video surveillance.
%GigaSight \cite{simoens2013scalable} blacks out sensitive information from video frames.
But in a photographing scenario,
 the challenge is how to match people's privacy requests with people in the photo.
Face recognition is a alternative way to solve the matching problem, \eg Eigenfaces \cite{turk1991eigenfaces}.
% and Fisherfaces\cite{belhumeur1997eigenfaces}.
%Many face descriptors are proposed for face recognition, \eg Eigenfaces \cite{turk1991eigenfaces} and Fisherfaces\cite{belhumeur1997eigenfaces}.
%proposes a principal components based method to detect faces and recognize the faces of known individuals.
%Eigenface method is expected to suffer under variation in lighting direction.
%Fisherfaces\cite{belhumeur1997eigenfaces} is a method based on Fisher¡¯s Linear Discriminant,
% which produces well separated classes in a low-dimensional subspace.
% even under severe variation in lighting and facial expressions.
To use face recognition approaches, it requires the people to face to cameras.
Besides, during the information exchange process, face descriptors could be leaked to adversaries.
There are some work providing privacy-preserving face recognition leverages homomorphic encryption,
 by which a client can privately search for a specific face image in the face image database of a server, \eg, \cite{sadeghi2010efficient}.
%\cite{erkin2009privacy} leverages homomorphic encryption
% to recognize a face in a database of $M$ faces.
%It requires $O(\log M)$ rounds and is computationally expensive.
%\cite{sadeghi2010efficient} improves the scheme with cryptographic building blocks
% combing homomorphic encryption with garbled circuits,
% requiring only $O(1)$ rounds with smaller communication and computation cost.
Those methods provide privacy protection to the requested images
 as well as the outcome of the matching algorithm,
 but the computation overhead is large and the result is not secure against the service provider.

\CUTXY{
\paragraph{Image Segmentation and Representation.}
Many efforts have been devoted to image segmentation and content representation \cite{liu2007survey}.
Vijayanarasimhan et al. \cite{vijayanarasimhan2011efficient} proposes a branch-and-cut strategy for region-based object detection. They form the problem as a prize-collecting Steiner tree problem.
Doulamis et al. \cite{doulamisrepresentation} present a hierarchical image content representation approach.}

\paragraph{Graph matching.}
%Graph matching techniques are related to the people matching in our approach.
Graph matching methods have been applied in many tasks such as face recognition \cite{wiskott1997face}, fingerprint identification \cite{isenor1986fingerprint}
 and others \cite{hu2013graph}.
%schema matching \cite{melnik2002similarity}, malicious software classification \cite{park2010fast}, and the like.
Their application-oriented specifications,
 \eg, assumptions about node numbers, graph structure and pre-knowledge of correspondences,
 make them difficult to be applied in this work.
%They target different applications with varying problem specifications and constraints.
%Laurenz et al. \cite{wiskott1997face} leverage fiducial points on human faces to facilitate the matching in which a set of individual model graphs are combined into a stack-like structure, called a face bunch.
%The Graduated Assignment Algorithm \cite{gold1996graduated} targets different matching problems such as the edge weighted graph matching, attributed relational graph matching, and the like.
%Some recent efforts \cite{caetano2009learning}\cite{leordeanu2012unsupervised} introduce the machine learning concepts into graph matching in which the train step uses pairs of graphs with fully correct correspondences.
%The recent work of Hu et al. \cite{hu2013graph} propose a matching scheme which leverages partially known correspondences as anchors.

\CUTXY{
\paragraph{Secret Exchange}
Diffie-Hellman key exchange is a well known protocol proposed to
distribute a session key between two parties through an untrusted channel.
Over the years, several papers have attempted to extend the well-known Diffie-Hellman
key exchange to the multi-party setting \cite{burmester1995secure, ateniese2000new,bresson2002group}.
Dynamic group Diffie-Hellman protocols for authenticated key exchange are designed to work in a scenario where a group of parties want to join and leave the multicast group at any given time\cite{bresson2001provably}.}

\paragraph{Privacy-Preserving Distance Computation.}
%For the random perturbation-based algorithms,
% the original data distributions can be reconstructed
% with some fair degree of accuracy, but mutual Euclidean
% distances between individual data points are not preserved.
Euclidean distance can be computed privately among parties
 using secure multi-party computation (SMC) methods \cite{katz2008predicate, sadeghi2010efficient} or
garble circuit \cite{sadeghi2010efficient}.
However,
 they usually require online interactions among data owners.
Moreover, their large computation cost and ciphertext size make them unsuitable for mobile applications.
\cite{mukherjee2006privacy} proposes an approach
 using Fourier-related transforms
 to hide accurate data values and to approximately preserve Euclidean
 distances among them.
It works well for some data mining purpose on large datasets,
 but the transformation is public and deterministic and it cannot prevent malicious user from dictionary attack.

\vspace{-0.08in}
\section{Conclusion}
\vspace{-0.08in}
\label{sec:conclusion}
In this work, we present a new approach InvisibleMe to protect users' portrait privacy during photo taking and sharing.
With our system, users that are unwilling to be photographed will be automatically erased from the pictures in a lightweight and privacy-preserving way.
To achieve this goal, we propose the integrated system model, a graph
matching scheme to locate people in pictures and a privacy-preserving
vector distance computation method.
We have fully implemented our protocol,
and thoroughly evaluated our design.
%the protection performance and the  overhead of the system.
%In future work, we plan to adapt the technique to other application scenarios.

\vspace{-0.08in}
{\small

\bibliographystyle{IEEEtran}
\bibliography{ref}

\begin{thebibliography}{10}
\providecommand{\url}[1]{#1}
\csname url@samestyle\endcsname
\providecommand{\newblock}{\relax}
\providecommand{\bibinfo}[2]{#2}
\providecommand{\BIBentrySTDinterwordspacing}{\spaceskip=0pt\relax}
\providecommand{\BIBentryALTinterwordstretchfactor}{4}
\providecommand{\BIBentryALTinterwordspacing}{\spaceskip=\fontdimen2\font plus
\BIBentryALTinterwordstretchfactor\fontdimen3\font minus
  \fontdimen4\font\relax}
\providecommand{\BIBforeignlanguage}[2]{{%
\expandafter\ifx\csname l@#1\endcsname\relax
\typeout{** WARNING: IEEEtran.bst: No hyphenation pattern has been}%
\typeout{** loaded for the language `#1'. Using the pattern for}%
\typeout{** the default language instead.}%
\else
\language=\csname l@#1\endcsname
\fi
#2}}
\providecommand{\BIBdecl}{\relax}
\BIBdecl

\bibitem{banglass}
``Business insider,
  http://www.businessinsider.com/seattle-bar-bans-google-glass-2013-3.''

\bibitem{google-law}
``http://www.franken.senate.gov/?p=press\_release\&id=2699.''

\bibitem{luo2007person}
J.~Luo, Y.~Ma, E.~Takikawa, S.~Lao, M.~Kawade, and B.-L. Lu, ``Person-specific
  sift features for face recognition,'' in \emph{ICASSP}.\hskip 1em plus 0.5em
  minus 0.4em\relax IEEE, 2007.

\bibitem{turk1991eigenfaces}
M.~Turk and A.~Pentland, ``Eigenfaces for recognition,'' \emph{Journal of
  cognitive neuroscience}, vol.~3, no.~1, pp. 71--86, 1991.

\bibitem{katz2008predicate}
J.~Katz, A.~Sahai, and B.~Waters, ``Predicate encryption supporting
  disjunctions, polynomial equations, and inner products,'' in
  \emph{EUROCRYPT}, 2008, pp. 146--162.

\bibitem{sadeghi2010efficient}
A.-R. Sadeghi, T.~Schneider, and I.~Wehrenberg, ``Efficient privacy-preserving
  face recognition,'' in \emph{ICISC}, 2010, pp. 229--244.

\bibitem{leibe2005pedestrian}
B.~Leibe, E.~Seemann, and B.~Schiele, ``Pedestrian detection in crowded
  scenes,'' in \emph{CVPR}.\hskip 1em plus 0.5em minus 0.4em\relax IEEE, 2005.

\bibitem{simoens2013scalable}
P.~Simoens, Y.~Xiao, P.~Pillai, Z.~Chen, K.~Ha, and M.~Satyanarayanan,
  ``Scalable crowd-sourcing of video from mobile devices,'' in
  \emph{Mobisys}.\hskip 1em plus 0.5em minus 0.4em\relax ACM, 2013.

\bibitem{facereconstruction}
``Face reconstruction, www.cs.princeton.edu/~cdecoro/eigenfaces/.''

\bibitem{xu2009feeling}
T.~Xu and Y.~Cai, ``Feeling-based location privacy protection for
  location-based services,'' in \emph{CCS}.\hskip 1em plus 0.5em minus
  0.4em\relax ACM, 2009.

\bibitem{komodakis2006image}
N.~Komodakis, ``Image completion using global optimization,'' in
  \emph{CVPR}.\hskip 1em plus 0.5em minus 0.4em\relax IEEE, 2006.

\bibitem{criminisi2004region}
A.~Criminisi, P.~P{\'e}rez, and K.~Toyama, ``Region filling and object removal
  by exemplar-based image inpainting,'' \emph{IEEE Transactions on Image
  Processing}, vol.~13, no.~9, pp. 1200--1212, 2004.

\bibitem{viola2004robust}
P.~Viola and M.~J. Jones, ``Robust real-time face detection,''
  \emph{International journal of computer vision}, vol.~57, no.~2, pp.
  137--154, 2004.

\bibitem{deng2001unsupervised}
Y.~Deng and B.~Manjunath, ``Unsupervised segmentation of color-texture regions
  in images and video,'' \emph{IEEE TPAMI}.

\bibitem{wiskott1997face}
L.~Wiskott, J.-M. Fellous, N.~Kuiger, and C.~Von Der~Malsburg, ``Face
  recognition by elastic bunch graph matching,'' \emph{IEEE TPAMI}, vol.~19,
  no.~7, pp. 775--779, 1997.

\bibitem{isenor1986fingerprint}
D.~Isenor and S.~G. Zaky, ``Fingerprint identification using graph matching,''
  \emph{Pattern Recognition}, vol.~19, no.~2, pp. 113--122, 1986.

\bibitem{hu2013graph}
N.~Hu, R.~M. Rustamov, and L.~Guibas, ``Graph matching with anchor nodes: A
  learning approach,'' in \emph{CVPR}.\hskip 1em plus 0.5em minus 0.4em\relax
  IEEE, 2013.

\bibitem{kuhn1955hungarian}
H.~W. Kuhn, ``The hungarian method for the assignment problem,'' \emph{Naval
  research logistics quarterly}, vol.~2, no. 1-2, pp. 83--97, 1955.

\bibitem{datar2004locality}
M.~Datar, N.~Immorlica, P.~Indyk, and V.~S. Mirrokni, ``Locality-sensitive
  hashing scheme based on p-stable distributions,'' in \emph{Symposium on
  Computational Geometry}.\hskip 1em plus 0.5em minus 0.4em\relax ACM, 2004.

\bibitem{burmester1995secure}
M.~Burmester and Y.~Desmedt, ``A secure and efficient conference key
  distribution system,'' in \emph{EUROCRYPT'94}, pp. 275--286.

\bibitem{lee2006distributed}
P.~P. Lee, J.~C. Lui, and D.~K. Yau, ``Distributed collaborative key agreement
  and authentication protocols for dynamic peer groups,'' \emph{TON}, vol.~14,
  no.~2, pp. 263--276, 2006.

\bibitem{senior2005blinkering}
A.~Senior, S.~Pankanti, A.~Hampapur, L.~Brown, Y.-L. Tian, and A.~Ekin,
  ``Blinkering surveillance: Enabling video privacy through computer vision,''
  in \emph{Security \& Privacy}.\hskip 1em plus 0.5em minus 0.4em\relax IEEE,
  2005.

\bibitem{zhang2005hiding}
W.~Zhang, S.-C.~S. Cheung, and M.~Chen, ``Hiding privacy information in video
  surveillance system.'' in \emph{ICIP}, 2005.

\bibitem{mukherjee2006privacy}
S.~Mukherjee, Z.~Chen, and A.~Gangopadhyay, ``A privacy-preserving technique
  for euclidean distance-based mining algorithms using fourier-related
  transforms,'' \emph{The VLDB Journal}, 2006.

\end{thebibliography}

}

%\vspace{3in}
%\appendix
%\input{app.tex}

\end{document}